\documentclass[12pt,manuscript]{emulateapj}
\usepackage{natbib,graphics,graphicx}
\hfuzz=\maxdimen
\newcommand{\KAB}  {\hbox{$K_{\rm AB}$}}
\newcommand{\bsex}{\hbox{\tt SExtractor}}
\newcommand{\zph}  {\hbox{$z_{\rm p}$}}
\newcommand{\zsp}  {\hbox{$z_{\rm s}$}}

\newcommand{\mum}  {\hbox{$\mu{\rm m}$}}
\newcommand{\muJy}  {\hbox{$\mu{\rm Jy}$}}
\newcommand{\sbzks}{\hbox{sBzKs}}
\newcommand{\pbzks}{\hbox{pBzKs}}
\newcommand{\sgzks}{\hbox{sgzKs}}
\newcommand{\pgzks}{\hbox{pgzKs}}
\newcommand{\gzks}{\hbox{gzKs}}
\newcommand{\etal}{\hbox{et al.\,}}
\newcommand{\gsim}{\lower.5ex\hbox{$\; \buildrel > \over \sim \;$}}
\newcommand{\lsim}{\lower.5ex\hbox{$\; \buildrel < \over \sim \;$}}
\newcommand{\psim}{\lower.5ex\hbox{$\; \buildrel \propto \over \sim \;$}}
\newcommand{\Lsun}{\mbox{$L_\sun$}}
\newcommand{\Msun}{\mbox{$M_\sun$}}
\slugcomment{Accepted for publication in the Astrophysical Journal} 
\shorttitle{z$\sim$2 galaxies in the AEGIS}
\shortauthors{Fang \etal}
\begin{document}
%\title{Passive and Star-forming Galaxies at $1.4 \leq z \leq2.5$ 
%in the AEGIS Field}  
\title{PASSIVE AND STAR-FORMING GALAXIES AT $1.4 \leq z \leq2.5$
IN THE AEGIS FIELD}
\author{
G{\scriptsize UANWEN} F{\scriptsize ANG}\altaffilmark{1,2,3}, X{\scriptsize U} K{\scriptsize ONG}\altaffilmark{1,3,4}, 
Y{\scriptsize ANG} C{\scriptsize HEN}\altaffilmark{1}, {\scriptsize AND} X{\scriptsize UANBIN} L{\scriptsize IN}\altaffilmark{1,4}
}
\affil{}
\altaffiltext{1}{Center for Astrophysics, University of Science and
Technology of China, Hefei 230026, China; xkong@ustc.edu.cn, wen@mail.ustc.edu.cn}
\altaffiltext{2}{Harvard-Smithsonian Center for Astrophysics, 60 
Garden Street, Cambridge, MA 02138, USA}
\altaffiltext{3}{Key Laboratory for the Structure and Evolution of 
Celestial Objects, Chinese Academy of Sciences, Kunming 650011, China}
\altaffiltext{4}{Key Laboratory for Research in Galaxies and Cosmology,
   USTC, Chinese Academy of Sciences, Hefei 230026, China}
%\email{xkong@ustc.edu.cn, wen@mail.ustc.edu.cn}

%
\received{2011 April 21}
\accepted{2012 March 27}

\begin{abstract}
Using a simple two-color selection based on $g$-, $z$-, and $K$-band 
photometry, we pick out 1609 star-forming galaxies (\sgzks)
and 422 passively evolving galaxies (\pgzks) at $z\sim2$ from a $K$-band-
selected sample ($K_{\rm AB} < 22.0$) in an area of $\sim 0.44$ deg$^{2}$ 
of the All-wavelength Extended Groth Strip International Survey.  
The number counts of \pgzks\ in our sample turn over at $K_{\rm AB} \sim 21.0$, 
and both the number of faint and bright objects (including \sgzks\ and \pgzks)
exceed the predictions of a recent semi-analytic model of galaxy 
formation, a more successful model is need to explain this diversity. 
We also find that the star formation rate (SFR) and specific SFR (sSFR) of \sgzks\ 
increases with redshift at all masses, implying that star-forming galaxies
were much more active on average in the past. Moreover, the sSFR
of massive galaxies is lower at all redshifts, suggesting that star formation
contributes more to the mass growth of low-mass galaxies 
than to high-mass galaxies. From {\it Hubble Space Telescope} Wide Field Camera 3 near-infrared imaging data,
we find that morphologies of $z\sim2$ galaxies not only have diffuse structures 
with lower $G$ and higher $M_{20}$ values, but also have single-object 
morphologies (higher $G$ and lower $M_{20}$), implying that there are 
morphological variety and different formation process for these galaxies at $z\sim2$.
Finally, we also study the fraction of active galactic nuclei (AGNs) in the \gzks,
82 of 828 \gzks\ with four IRAC bands can be classified as AGNs ($\sim$ 10\%). 
Most of these AGN candidates have $L_{\rm 0.5-10\ keV}>10^{41}\,\rm erg\,s^{-1}$.
\end{abstract}


\section{Introduction}\label{sec:intro}

Understanding when and how the most massive galaxies in the universe 
formed is one of the most outstanding problems in cosmology and 
galaxy formation (Conselice \etal 2007). 
A number of observations suggest that the era of $z\sim2$ is 
important in galaxy formation and evolution for various reasons: the cosmic star 
formation rate (SFR) density begins to drop at $z\sim2$ from a flat 
plateau at higher redshifts; the morphological type mix of field 
galaxies changes remarkably at $z\sim2$; the number density of QSOs 
has a peak at $z\sim2$; $\sim50\%-70\%$ of the stellar mass 
assembly of galaxies took place in the redshift range $1<z<3$ 
(Dickinson \etal 2003; Fontana \etal 2003; Steidel \etal 2004; 
Kong \etal 2006; Richards \etal 2006; Arnouts \etal 2007; 
Pozzetti \etal 2007; Noeske \etal 2007). However, in the redshift 
range $1.4<z<3.0$, identifiable spectral features such as \ion{Ca}{2} H\&K 
and the 4000 \AA{} break (for passive galaxies) and [\ion{O}{2}]$\lambda3727$, 
[\ion{O}{3}]$\lambda5007$, H$\alpha$,\ and H$\beta$\ (for star-forming ones) 
move out of the optical bands, thus near-infrared imaging and spectroscopy 
become essential. On the other hand, near-infrared galaxy samples offer several 
advantages compared to purely optical selections (Cowie \etal 1994). 
They allow us to select galaxies at $z>1$ in the rest-frame optical, 
correspond more closely to a stellar-mass-selected sample, and are 
less prone to dust extinction.

There are many ways of using optical and infrared (IR) colors
to select the sample of galaxies at $z\sim2$. 
A sample of IR-luminous dust-obscured galaxies
(DOGs), with a median $\nu L_{\nu}(8\micron)\sim4\times10^{11}~\Lsun$,
selected to have very red $R - [24]$ color, was recently presented by
Dey \etal (2008). From spectroscopic observations DOGs were found to
have a tight redshift distribution around $z\sim2$, very similar to
that of the submillimeter-selected galaxies (Chapman \etal 2005).
Another different sample, based on near-infrared color selection,
is that distant red galaxies with $(J-K)_{\rm vega}>2.3$ (DRGs; Franx 
\etal 2003; van Dokkum \etal 2004). These sources are redder and strong 
clustering and are believed to be more massive than $10^{11}~\Msun$ 
(Labbe \etal 2005; Papovich \etal 2006; van Dokkum \etal 2006; 
Quadri \etal 2008). A third sample uses {\it Spitzer}/IRAC color to
select galaxies at $z>1.4$. In the redshift range $1.4<z<2.7$, the
four IRAC bands probe the rest-frame near-infrared bands where
galaxy spectral energy distributions (SEDs) have similar shape, thus
the IRAC colors are very robust in determining redshift in this
range (Farrah \etal 2008; Desai \etal 2009; Huang \etal 2009; 
Fadda \etal 2010; Fiolet \etal 2010; Fang \etal 2011).
Most sources of these samples are starburst-dominated ultraluminous
infrared galaxies (ULIRGs; $L_{\rm IR}>10^{12}~\Lsun$), with an
apparent 1.6 \micron\ stellar bump in the IRAC channels.

An independent way to select galaxies at $z\sim2$ is to use $BzK$
colors. Daddi \etal (2004) introduced a new photometric 
technique, $BzK=(z-K)_{\rm AB}-(B-z)_{\rm AB}=-0.2$ and 
$(z-K)_{\rm AB}=2.5$, for obtaining a virtually complete sample of 
star-forming ($BzK\geq-0.2$) and passively evolving galaxies 
($BzK<-0.2$ and $(z-K)_{\rm AB}>2.5$) at $1.4\leq z \leq2.5$. 
These criteria are reddening independent for star-forming galaxies 
in the selected redshift range, thus can be used to select the 
reddest dust-extinguished galaxies. This should allow a relatively 
unbiased selection of star-forming (\sbzks) and passively evolving 
galaxies (\pbzks) at $z\sim2$ within the magnitude limit of the 
sample studied. Based on $BRIzJK$ photometry from Subaru and New Technology Telescope (NTT) 
over two separate fields, using the $BzK$ technique, Kong \etal (2006) 
have obtained a sample of $\sim500$ \sbzks\ and $\sim160$ \pbzks, 
which were identified over an area of $\sim 920$ arcmin$^2$ 
to $\KAB=20.8$, of which 320 arcmin$^2$ are complete to $\KAB=21.8$. 
They found that the log of the number counts of \pbzks\ flattens out 
by $\KAB \sim 20.8$. Similar conclusions were also found by Lane \etal (2007) 
and McCracken \etal (2010), with much larger samples.

Investigations of the physical properties (i.e., stellar mass, 
luminosity, size, clustering, SFR, 
morphology, etc.) of various types 
of galaxies from high to low redshift are required for better 
understanding the history of the galaxy formation and evolution. 
Dunne \etal (2009) studied the star formation history
of {\it K}-selected galaxies in the UKIDSS Ultra-Deep Survey. They 
found that the specific star formation rate (sSFR) for 
{\it K}-selected sources rises strongly with redshift at all stellar
masses. In the same field, Williams \etal (2009, 2010) investigated
the physical properties (including clustering, size, stellar mass,
surface density, SFR, color, etc.) of quiescent and star-forming
galaxy populations to $z = $0--2 with pure photometric data,
employing a novel rest-frame $U-V$ versus $V-J$ technique 
(Wuyts \etal 2007; Wolf \etal 2009; Balogh \etal 2009;
Whitaker \etal 2010; Patel \etal 2011). As shown in
Figure 8 of Williams \etal (2009) and Figure 2 of Brammer
\etal (2011), the red quiescent and star-forming galaxies are 
found to occupy two distinct population in the rest-frame 
$U-V$ versus $V-J$ color space. This bimodal behavior is still
seen up to $z\sim2$. At all redshifts, furthermore, massive 
quiescent galaxies occupy the extreme high end of the surface
density distribution and a tight mass--size correlation, while
star-forming show a broad range of both densities and sizes.
On the other hand, the sizes and surface densities of massive
quiescent and star-forming galaxies evolve as simple power-law
behavior in ($1+z$), with more massive galaxies exhibiting
faster evolution. In the meantime, Brammer \etal (2009, 2011),
van Dokkum \etal (2010), and Whitaker \etal (2011) 
provided further evidence for the results above, using the
sample from the NEWFIRM Medium Band Survey (NMBS; van Dokkum 
\etal 2009). More studies for the {\it K}-selected galaxies
have been performed up to $z=5$ (Cimatti \etal 2002;
Abraham \etal 2004; Drory \etal 2005, 2009; Fontana
\etal 2006; Pozzetti \etal 2007; P{\'e}rez-Gonz{\'a}lez \etal 2008; 
Furusawa \etal 2011).

Morphologies are also essential in studying galaxy mass assembly 
history and evolution. For the morphologies of galaxies at $z\sim2$, 
the traditional Hubble sequence of regular spirals and elliptical 
galaxies has not settled into place by $z\sim2$, and a much higher 
frequency of diffuse, clumpy, and irregular structures is observed 
among star-forming systems (Lotz \etal 2006; Szomoru \etal 2011). 
With the advent of {\it Hubble Space Telescope} {\it HST}/Wide Field Camera 3 (WFC3), it is now possible 
to obtain a complete rest-frame optical morphological census for
massive galaxies at $z\sim2$. 

Understanding the nature of active galactic nuclei (AGNs) and the 
galaxies that host them is important for such diverse goals as 
pinpointing the sources of the cosmic X-ray and IR backgrounds 
and deriving the star formation history of the universe.
Multi-wavelength surveys are particularly important for
the study of AGNs because their appearance in different
wavelength regimes can be quite different.
In this work, we will discuss the fraction of AGNs in our sample
by using the IRAC band color criteria of Stern \etal 
(2005) and mid-IR spectral index (Barmby \etal2006; Park \etal 2010).

To study the physical properties of galaxies in the redshift 
range $1.4\lsim z \lsim 2.5$, in this paper, we select a sample 
of passively evolving and star-forming galaxies to $\KAB<22$ in the 
All-wavelength Extended Groth Strip International Survey (AEGIS) data set.
The paper is organized as follows. We describe the multi-band spatial- 
and ground-based observations of the AEGIS field, and introduce data 
reduction, photometric redshifts, and method for galaxy sample 
selection in Section 2. Section 3 presents source number counts 
(\sgzks\ and \pgzks). Section 4 presents the physical 
properties of \sgzks\ (including the estimate SFR and stellar 
mass of star-forming galaxies and the stellar mass versus SFR 
relation). Section 5 describes the morphologies of both \sgzks\ and \pgzks\
in the AEGIS. Section 6 discusses the fraction of AGNs in our 
galaxy sample. Finally, a brief summary is presented in Section 7. 
All magnitudes and colors are in the AB system unless stated otherwise. Throughout 
the paper we adopt the following cosmology: 
$h = H_0$[km s$^{-1}$ Mpc$^{-1}$]$/100=0.71$, $\Omega_\Lambda =0.73$, 
$\Omega_M = 0.27$.

\section{Data and Sample Selection}\label{sec:sam}

The AEGIS is a collaborative effort to obtain both deep imaging 
covering all major wavebands from X-ray to radio and optical 
spectroscopy over a large area of sky ($0.5-1.0$\,deg$^2$) with the 
aim of studying the panchromatic properties of galaxies over the last half 
of the Hubble time. 
AEGIS is targeted on a special area of the sky, called the Extended 
Groth Strip (EGS; centered at $\alpha$(J2000)$ = 14^{\rm h}19^{\rm m}12^{\rm s}$,  
$\delta$(J2000)$=52^{\circ}43^{\prime}42^{\prime \prime}$), which has 
low-extinction, low-Galactic, and zodiacal infrared emission, has good 
schedulability by space-based observations, and has therefore 
attracted a wide range of deep observations at essentially every 
accessible wavelength over this comparatively wide field (Davis \etal 
2007).

\subsection{Data Observations}\label{sec:obs}
The EGS is a public deep multi-wavelength survey; this field provides 
an unique combination of area and depth at almost every waveband 
observable. In this paper, {\it Chandra}/ACIS X-ray, CFHT/MegaCam Legacy 
Survey optical, CFHT/CFH12K optical, Palomar/WIRC near-infrared, 
{\it Spitzer}/IRAC mid-infrared, and {\it Spitzer}/MIPS far-infrared imaging data 
or catalog of the EGS is used. Davis \etal (2007) provided more details 
on the data reduction and its performance.

For our analysis, specially, the {\it Chandra}/ACIS DR2 catalog presented 
by Laird et al.\,(2009) is used. The X-ray data reduction was performed 
using the CIAO data analysis software version 3.3. 
Basic data reduction proceeded in a manner similar to that described 
by Nandra \etal(2005). The vast majority of X-ray sources detected in 
the EGS field are AGNs. In the meantime, the EGS is also one of 
four $1.0\,$deg$^2$ fields covered by the CFHTLS deep survey 
(labeled CFHTLS-D3). The data covers the observed wavelength range 
3500\AA$<\lambda<$9400\AA\, in the $u^*$, $g'$, $r'$, $i'$, $z'$ 
filters (hereafter labeled $u$, $g$, $r$, $i$, $z$).
The limiting AB magnitudes is $u\sim27.0$, $g\sim28.3$, $r\sim27.5$, 
$i\sim27.0$, $z\sim26.4$ in the current data set (corresponding to the 
magnitude limit at $5\sigma$ point-source detection). 
Near-infrared data catalog in $J$ and $K$ bands are available over 
$\sim0.67$ deg$^2$ ({\it K}-band) with the $5\sigma$ magnitudes limits of 
$J_{\rm AB}\sim23.9$ and $K_{\rm AB}\sim22.5$ (Bundy \etal 2006).
Deep observations of the EGS with {\it Spitzer}/Infrared Array Camera (IRAC) 
cover an area of $\sim0.33$ deg$^2$ to a 50\% completeness limit of 
$1.5\,\muJy$ at $3.6\,\mum$ (Barmby \etal 2008). 
The catalog comprises 57,434 objects detected at $3.6\,\mum$, with 
84\%, 28\%, and 24\% also detected at $4.5\,\mum$, $5.8\,\mum$, 
$8.0\,\mum$. As for far-infrared data sets, we also use the {\it Spitzer}/Multiband Imaging Photometer for {\it Spitzer} 
(MIPS) $24\,\mum$ imaging data 
with the $5\sigma$ flux limit of $\sim77\,\muJy$ over $0.534$ deg$^2$. 
Sources are identified and photometry extracted with point-spread function (PSF) fitting using %PSF fitting using 
the {\tt DAOPHOT} software (Stetson 1987).

A catalog of objects with redshifts known from spectroscopic 
observations is very important to the construction of a photometric 
redshift code. For this reason, we use the DEEP2 (Deep Extragalactic 
Evolutionary Probe) spectroscopic data  obtained with the Deep Imaging 
Multi-Object Spectrograph boarded on the Keck 10 m telescope. 
The DEEP2-DR3 redshift catalog structures contain 24 tags for each 
object with 50,386 entries in the entire catalog. 
For the EGS field (DEEP2-field1), we only consider objects with 
redshift quality greater than or equal to 3. To specifically study
the accuracy of the photometric redshift, we also adopt spectroscopic
redshifts drawn from the Lyman break galaxy (LBG; $z\sim3$) catalog of 
Steidel \etal (2003) and from a small sample of ULIRGs ($z\sim2$)
presented in Huang \etal (2009) and Fang \etal (2011). 

The NMBS employs a new technique of using medium-bandwidth near-infrared 
filters to sample the Balmer/4000\AA\ break from $1.5<z<3.5$ at a higher 
resolution than the standard broadband near-infrared filters 
(van Dokkum \etal 2009), thereby improving the accuracy of photometric 
redshifts. A custom set of five medium-bandwidth filters in the wavelength range of
$1-1.8~\mum$ were fabricated for the NEWFIRM camera on the Mayall
4 m telescope on Kitt Peak for the NMBS (see Figure 1 of Whitaker \etal 2011).
The survey targets two fields within the COSMOS (Scoville \etal 2007) and 
AEGIS (Davis \etal 2007) surveys, chosen to take advantage of the wealth of 
publicly available ancillary data over a broad wavelength range.  
The more details of the observations, data reduction, and photometry
for the survey can be found in  Whitaker \etal (2011). The NMBS catalogs 
contain $\sim13,000$ galaxies at $z>1.5$ with accurate photometric 
redshifts and rest-frame colors. In this paper, we will use their
photometric redshifts in the AEGIS field.
   
\subsection{Data Reduction}\label{sec:data}

Source extraction for the CFHTLS optical images is performed with 
\bsex\, (Bertin \& Arnouts 1996) in the dual image mode ($i$ band is 
used as the reference image); a 2$''$diameter aperture is used for 
aperture magnitudes. The edges of the resulting final images, where 
the signal to noise of CFHTLS/MegaCam images is very low, are trimmed. 
We finally cross-correlate the $i$-band catalog (that included $u$, 
$g$, $r$, and $z$ photometry) with the $K$-band catalog, with a 
matching radius 2$''$. The final overlapping area of all our data sets is 
$\sim\,0.44$\,deg$^2$, and 16,228 objects are included in our final 
catalog ($\KAB\,<\, 22$ mag). Comparing to the NMBS in the AEGIS field 
($\sim\,0.21$\,deg$^2$), we have larger area and more faint galaxies 
(see Section 2.4). Stellar object are isolated with the color 
criterion $(z-K)_{\rm AB}\,<\,0.4(g-z)_{\rm AB}-0.45$, similar as Daddi \etal 
(2004). A total of 1847 objects are classified as stars, and the number of the final 
galaxy sample is 14,381. All magnitudes are corrected for Galactic extinction.

\subsection{Photometric Redshift}\label{sec:zphot}

\begin{figure*}
\centering
\includegraphics[angle=-90,width=0.95\textwidth]{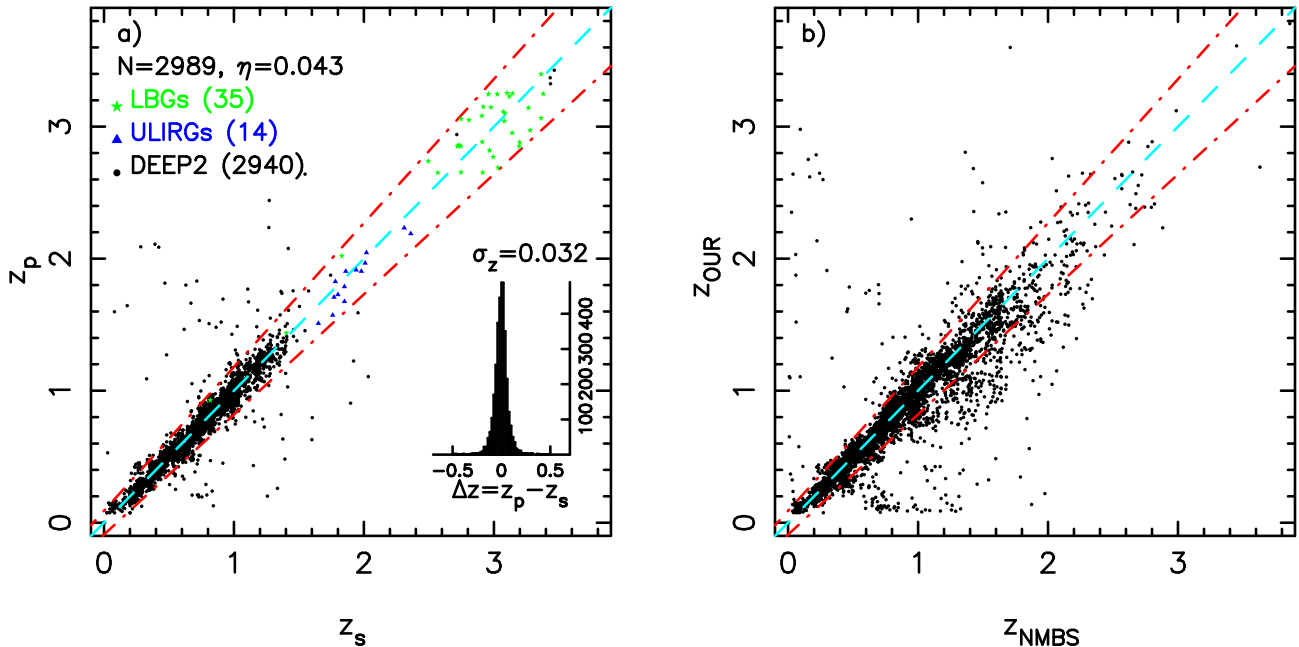}
\caption{(a) Comparison between spectroscopic redshift ($\zsp$) and photometric 
redshift ($\zph$). The dot-dashed lines are for $\zph=\zsp\pm0.10(1+\zsp)$. 
The number of galaxies with spectroscopic redshift and the fraction of 
catastrophic failures (($\zph-\zsp)/(1+\zsp)>0.1$) are listed in 
the top-left corner. The inset shows the $\Delta z=\zph-\zsp$ 
distribution, and the 1$\sigma$ dispersion is 0.032. The dashed 
line is $\zph=\zsp$. Filled triangles and stars represent ULIRGs 
at $z\sim2$ and LBGs at $z\sim3$ respectively. Filled circles
are drawn from the DEEP2-DR3 redshift catalog.
(b) Comparison between $z_{\rm NMBS}$ and $z_{\rm OUR}$. The $z_{\rm NMBS}$ 
are drawn from a public {\it K}-selected photometric catalog with  
photometric redshifts in the NMBS/AEGIS field (Whitaker \etal 2011). 
A total of 5533 galaxies are located in the overlapping area between NMBS and AEGIS.
\label{fig:zphot}}
\end{figure*}

An SED fitting technique based on an updated version of the {\tt HyperZ} 
code (Bolzonella et al. 2000) is used to calculate the photometric 
redshifts (hereafter $\zph$) for all the $\KAB<22$ galaxies in the EGS. 
A set of templates generated by Brammer \& van Dokkum (2007) for their 
{\tt EAZY} photometric redshift code are used. 
For each object, the SED derived from the 
observed magnitudes is compared to each template spectrum in turn, and 
the weighted mean redshift, computed in the confidence intervals at 
99\% around the main solution, from {\tt HyperZ}, is calculated.

To check the accuracy of the photometric redshift we derived, we first
compare the $\zph$ with the spectroscopic redshifts.
For the 14,381 galaxies in our sample, 2940 of them have spectroscopic
redshifts ($\zsp$) from the DEEP2-DR3 redshift catalog, which have 
a confidence level greater or equal to 97\%. In the meantime, spectroscopic 
redshifts are also available for 35 LBGs at $z\sim3$ 
and 14 ULIRGs at $z\sim2$ within the EGS field (14 ULIRGs presented in 
Huang \etal 2009 and Fang \etal 2011, 35 LBGs drawn from the catalog 
of Steidel \etal 2003). In Figure~\ref{fig:zphot}(a), we show
a comparison of the photometric redshift with the spectroscopic 
redshift for those 2989 galaxies. Filled triangles and stars correspond
to ULIRGs and LBGs with spectroscopic redshifts, respectively. From this 
figure, we find that our photometric redshifts are in good agreement 
with the spectroscopic redshifts, with an average 
$(\zph-\zsp)/(1+\zsp)=-0.014$. Only 128 of 2989 galaxies have 
$(\zph-\zsp)/(1+\zsp)>0.1$; the percentage of catastrophic failures
is about 4.3\%. The normalized median absolute deviation, $\sigma_{\rm NMAD}$, 
for our galaxy sample is $\sigma_{\rm NMAD}=0.032$. 

On the other hand, Whitaker \etal (2011) described the full details of 
the observations, data reduction, and photometry from the NMBS in the AEGIS field 
($\sim\,0.21$\,deg$^2$), also presented a public {\it K}-selected photometric 
catalog along with accurate photometric redshifts and rest-frame colors.
For this catalog, we will use their photometric redshifts ($z_{\rm NMBS}$).
We finally cross-correlate our catalog with the NMBS catalog in the AEGIS field, 
with a matching radius 1$''$. A total of 5533 galaxies are located in the overlapping 
area between NMBS and AEGIS. In Figure~\ref{fig:zphot}(b), 
our photometric redshifts ($z_{\rm OUR}$) are in good agreement
with NMBS photometric redshifts ($z_{\rm NMBS}$) in the AEGIS field, 
although there are some large dispersion.

\subsection{$z\sim2$ Galaxy Selection}\label{sec:samp}

\begin{figure*}
\centering
\includegraphics[angle=-90,width=0.95\textwidth]{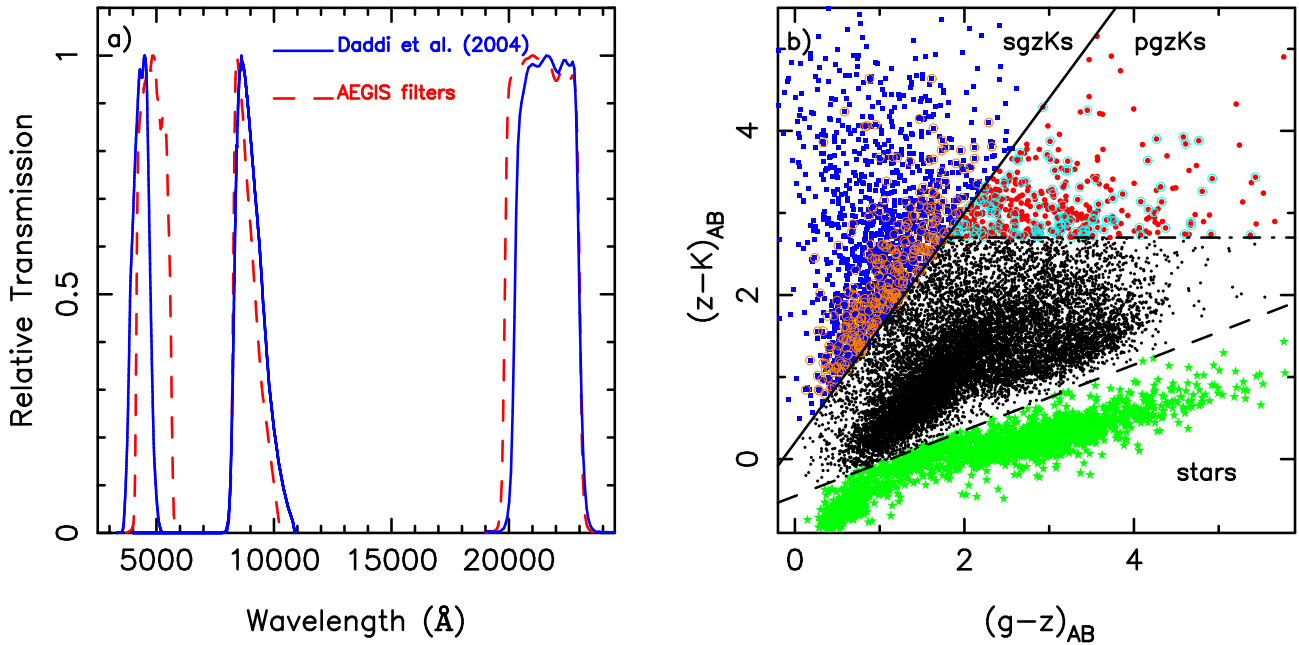}
\caption{
(a) Filter transmission profiles for the AEGIS and those used by Daddi \etal (2004). 
These profiles are normalized to a maximum throughput of 1. 
The solid lines correspond to the $B$-, $z$-, and $K_s$-band filter 
curves used in Daddi \etal (2004), respectively; the dashed lines 
correspond to the $g$-, $z$-, and $K$-band curves used in the AEGIS.
(b) Two-color $(z-K)_{\rm AB}$ vs. $(g-z)_{\rm AB}$ diagram for the 
galaxies in the EGS field. Galaxies at high redshifts are highlighted. 
The diagonal solid line defines the region 
$gzK = (z-K)_{\rm AB}-1.4(g-z)_{\rm AB}\geq0.2$ that is efficient 
to isolate $z>1.4$ star-forming galaxies (\sgzks). 
The horizontal dot-dashed line further defines the region 
$(z-K)_{\rm AB}>2.7$ that contains old galaxies at $z>1.4$ (\pgzks). 
The dashed line defines the region occupied by stellar objects. 
Filled stars show stellar objects in our fields with 
$(z-K)_{\rm AB} - 0.4(g-z)_{\rm AB}<-0.45$; filled squares represent \sgzks; 
filled circles represent \pgzks. The sample of gzKs (\pgzks\ and \sgzks)
with open circles from the overlapping area between NMBS and AEGIS. 
These gzKs mostly lie along the boundary with non-gzK objects; 
comparing to the photometric redshift distribution for gzKs in our sample, 
their photometric redshifts have lower values.
\label{fig:samp}}
\end{figure*}

To select a catalog of objects at $z\sim2$, we use the $BzK$ selection 
technique which was introduced by Daddi \etal (2004). This technique has 
been adopted and tested in several subsequent studies 
(e.g., Kong \etal 2006; Lane \etal 2007; Blanc \etal 2008; Dunne \etal 2009; 
McCracken \etal 2010; Onodera \etal 2010; Yoshikawa \etal 2010). 
To make the comparison possible with previous studies, we wanted our 
photometric selection criterion to match as closely as possible as the 
original $BzK$ selection introduced in Daddi \etal (2004).
Figure~\ref{fig:samp}(a) shows the transmission profiles of the VLT-$BzK$ 
filters, which were used in Daddi \etal (2004), the transmission 
profiles of the CFHT-$gz$, and the Palomar-$K$ filters of AEGIS, which are 
used by this paper. 
From this figure, we find that the shapes of those filters are different. 
To account for the differences, we carefully compare the stellar 
sequence in the AEGIS to that of Daddi \etal (2004), with the similar 
procedure outlined in Kong \etal (2006), and apply small color terms 
to $B-z$ and $g-z$, $z-K$ in AEGIS and in Daddi \etal (2004), in order 
to obtain a fully consistent match. 

From this analysis, we obtain our new color criteria $g-z$ and $z-K$ to 
select $z\sim2$ galaxies: $gzK = (z-K)_{\rm AB}-1.4(g-z)_{\rm AB} \geq 
0.2$ can be used to select $z>1.4$ star-forming galaxies (hereafter 
named \sgzks), $gzK<0.2$ and $(z-K)_{\rm AB}>2.7$ can be used to 
isolate $z>1.4$ passively evolving galaxies (hereafter named \pgzks).
To further elucidate the validity of these $gzK$ criteria, we generate
a set of templates, with different star formation history, by Bruzual 
\& Charlot (2003) stellar population synthesis models, and find that
galaxies at $1.4\lsim z\lsim 2.5$ with ongoing star formation indeed lie in the 
$gzK\geq0.2$ region as expected, passively evolving 
galaxies lie in the $gzK<0.2$ and $(z-K)_{\rm AB}>2.7$ region of the 
$g-z$ and $z-K$ diagram.

Figure~\ref{fig:samp}(b) shows the $gzK$ color diagram of $K$-selected objects 
in the AEGIS. Using the color criterion $gzK = (z-K)_{\rm AB}-1.4(g-z)_{\rm AB} 
\geq 0.2$ (the black solid line), 1609 galaxies with $\KAB <22$ are 
selected as \sgzks\, (blue squares), which occupy a narrow range on 
the left of the solid line in Figure~\ref{fig:samp}(b).  
Using $gzK<0.2$ and $(z-K)_{\rm AB}>2.7$ (dot-dashed line), 422 
objects are selected as candidate \pgzks\, (red circles), which 
lie in the top-right part of Figure~\ref{fig:samp}(b). 
The sample of gzKs (\pgzks\ and \sgzks) with 
open circles from the overlapping area between NMBS and AEGIS. These gzKs
mostly lie along the boundary with non-gzK objects, comparing to
the photometric redshift distribution for gzKs in our sample, their 
photometric redshifts have lower values.
The surface density of $\KAB<22$  $\sgzks+\pgzks$ is $\sim 1.28$ 
arcmin$^{-2}$, which is in good agreement with those
presented in Daddi \etal (2004), about $\sim1.1$ arcmin$^{-2}$, 
and Kong \etal (2006), about $\sim 1.4$ arcmin$^{-2}$.

\begin{figure*}
\centering
\includegraphics[angle=-90,width=0.95\textwidth]{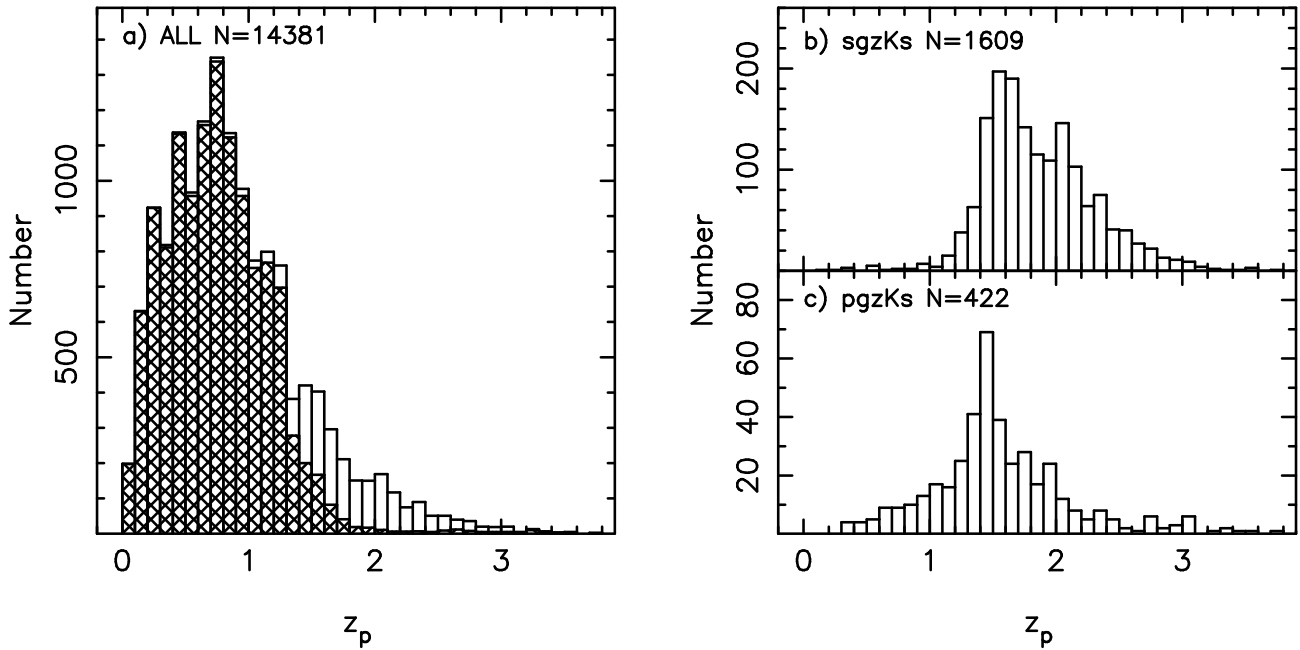}
\caption{
Photometric redshift distributions for the $K$-selected galaxies in 
our sample. The left panel shows the redshift distributions for 
all galaxies and all non-gzKs galaxies (cross-hatched). 
The top-right panel shows the redshift distributions of our \sgzks\ 
sample (1609 galaxies), the bottom-right panel shows the redshift 
distributions of \pgzks\ (422 galaxies). 
\label{fig:reds}}
\end{figure*}

In Figure~\ref{fig:reds}(a), we show the histogram of the photometric 
redshift for all $K$-selected galaxies in our sample. To check how 
efficient is our $gzK$ color criteria in singling out $z>1.4$ 
galaxies, the cross-hatched region shows all non-gzK galaxies
in our sample. From this figure, we find that only 4\% of galaxies 
with photometric redshift $z>1.4$ cannot be selected by the $gzK$ 
color criteria, which means the $gzK$ criteria are quite efficient 
in singling out high-redshift galaxies. Figure~\ref{fig:reds}(b) shows 
the distribution of the photometric redshifts of \sgzks, with a median 
value $z\sim1.8$. Most of them have redshifts at the range $1.4\leq z\leq2.5$, 
only $8.5\%$ of them with $z<1.4$ and $8.3\%$ with $z>2.5$. 
We show the redshift distribution of \pgzks\ in 
Figure~\ref{fig:reds}(c); it clearly shows that there is a considerable 
fraction (12.8\%) of objects at low redshift ($z<1.0$). The median 
redshift value of  \pgzks\ is $\sim1.5$, which is similar to
that ($\bar{z} \sim1.4$) of McCracken \etal (2010) . 

\section{Number Counts of gzKs}\label{sec:cou}

\begin{figure*}
\centering
\includegraphics[angle=-90,width=0.95\textwidth]{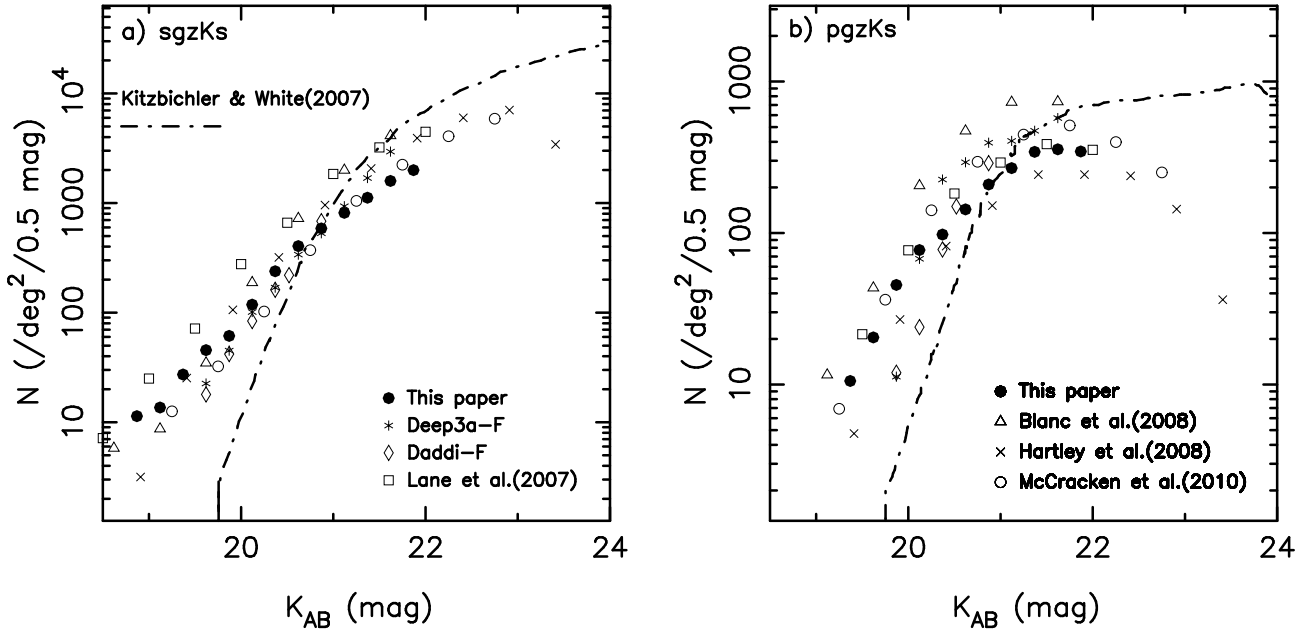}
\caption{
$K$-band differential number counts for star-forming galaxies 
(\sgzks) and passively evolving galaxies (\pgzks) at $z\sim2$.
For comparison, we have overplotted the number counts for $z\sim2$
galaxies from the literature and the predictions of model of 
Kitzbichler \& White (dot-dashed line). 
Left: \sgzks. Right: \pgzks. 
\label{fig:cou}}
\end{figure*}

Figure~\ref{fig:cou} shows the differential number counts of 
\sgzks\ (left panel) and \pgzks (right panel). 
In both panels of Figure.~\ref{fig:cou}, results from the Deep3a-F
(asterisks) and Daddi-F (open diamonds) from Kong \etal (2006), 
UKIDSS (open squares, Lane \etal 2007; crosses, Hartley \etal 2008),
MUSYC (open triangles, Blanc \etal 2008), and COSMOS (open circles, 
McCracken \etal 2010) are also plotted for comparison.
Figure~\ref{fig:cou}(a) shows the number counts of \sgzks, no correction
 for incompleteness is applied to our data. 
As shown in the figure, our number counts of \sgzks\ in the EGS field 
are in good agreement with the number counts of $z\sim2$ star-forming
galaxies in Kong \etal (2006), Blanc \etal (2008), and McCracken \etal 
(2010). However, the counts presented by Lane \etal (2007) are offset
compared to our counts, which is also found by McCracken \etal (2010), 
and the cause for this discrepancy may arises from an incorrect 
transformation to Daddi \etal's (2004) filter set in Lane \etal (2007). 

Figure~\ref{fig:cou}(b) shows the number counts of \pgzks. Our number 
counts of \pbzks, in general, are in good agreement with those 
presented in the literature for $z>1.4$ passively evolving galaxies.
At bright magnitudes (e.g., $\KAB<21.0$) our counts are in very good
agreement with the pBzKs counts from McCracken \etal (2010). 
However, the number counts of ours and those of McCracken \etal 
(2010) are slightly below those from Blanc \etal (2008) at 
faint magnitudes. Combining literature data with the present work, 
we can examine the global shape of the $z>1.4$ passively evolving 
galaxy counts over a very wide dynamic range. This strongly suggests 
a break feature in the slope at $\KAB<21.0$, which was found in 
Kong \etal (2006) already.

The dot-dashed lines in Figure~\ref{fig:cou}(a) and (b) 
show the number counts from the semi-analytic model of 
Kitzbichler \& White (2007). The number counts of quiescent galaxies 
and star-forming galaxies were derived by McCracken \etal (2010), 
following the approach of Daddi \etal (2007). From 
Figure~\ref{fig:cou}, we find the models overpredict the number 
of faint galaxies, but underpredict the number of bright galaxies 
in all two plots. Given the narrow redshift range of our $z>1.4$ 
galaxy population, apparent $K$ magnitude is a good proxy for 
absolute $K$ magnitude, which can itself be directly related to 
underlying stellar mass (Daddi \etal 2004). This implies that these 
models predict too many small low-mass star-forming and passively 
evolving galaxies, and too few large high-mass star-forming and 
passively evolving galaxies at $z>1.4$. Similar conclusions were 
drawn by Fontana \etal (2006) and McCracken \etal (2010). Therefore, 
models incorporating AGN feedback similar to Kitzbichler \& White 
underpredicted the number of high-mass galaxies, however predicted 
too many low-mass galaxies.

\section{Physical Properties of \sgzks}\label{sec:sgzk}

In this section, physical properties of \sgzks, such as  stellar mass 
($M_*$), SFR, and sSFR (SFR$/M_*$), are derived on the basis of the 
present photometric data.

\subsection{Stellar Mass Estimates}

\begin{figure*}
\centering
\includegraphics[angle=-90,width=0.95\textwidth]{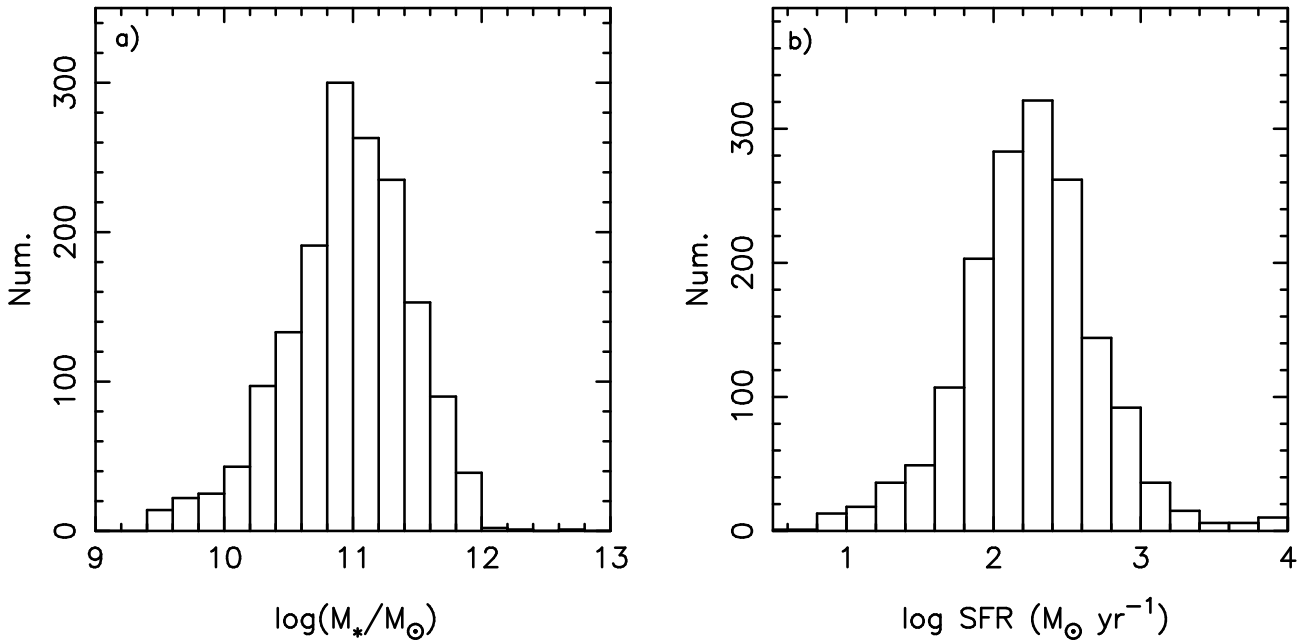}
\caption{
Stellar mass and star formation histogram of \sgzks\ in the EGS
field. The left panel (a) is the plot for stellar mass, and the right panel 
(b) is the plot for star formation rate.
\label{fig:sfr}}
\end{figure*}

In order to estimate the stellar mass ($M_*$) of \sgzks, we perform SED fitting
 of the multi-band photometry as described above with population synthesis
models. The SED templates are generated with the SPS package developed
 by Bruzual \& Charlot (2003). We assume an universal initial mass 
function (IMF) from Chabrier (2003). Masses derived assuming this IMF
can be converted to Salpeter by adding 0.3 dex (Bundy \etal 2005).
A grid of model templates spanning a range of star formation histories 
(parameterized as an exponential), ages, metallicities and dust content 
are used. The stellar mass-to-luminosity ($M/L$) ratios from the best-fit 
templates and the photometric redshifts (spectroscopic redshifts are
adopted if available) from Section 2.3  are used to calculate the 
stellar mass. The histogram for the stellar mass of the \sgzks\ is shown in 
Figure~\ref{fig:sfr}(a). About 50\% of the \sgzks\ in our sample 
have $M_*>10^{11} M_\odot$, and the median stellar mass of \sgzks\ is 
$\sim 8.8\times10^{10}M_\odot$.

\subsection{Star Formation Rates}

The ultraviolet continuum can be used to estimate the luminosity from
young stars and hence the SFR. We derive the SFR of each \sgzks\ based 
on its rest-frame 2800\AA\ luminosity, which is calculated from the same 
best-fitting SED template as used in the estimation of the photometric redshift.
Since our multi-band photometric data completely cover the rest-frame 
2800\AA\ for the redshift range ($1.0\lsim z \lsim3.0$) of \sgzks, the 
estimated rest-frame 2800\AA\ luminosity is a robust quantity, 
especially in the case where a galaxy is detected in both bands which 
straddle the rest-frame 2800\AA. The color excess, $E(B-V)$, from the SED 
fitting is used to derive the dust-corrected UV luminosity.
Then, we convert the dust-corrected rest-frame 2800\AA\ luminosities 
into SFRs using the calibration by Kennicutt (1998): 
${\rm SFR}\,(M_\odot\ {\rm yr}^{-1})=1.4\times10^{-28}\ 
L_\nu\,({\rm erg}\, {\rm s}^{-1}\, {\rm Hz}^{-1})$.

The SFR histogram of the \sgzks\ is shown in Figure~\ref{fig:sfr}(b). 
About 85\% of \sgzks\ in our sample ($\KAB < 22$) have 
SFR $>70\,M_\odot$\,yr$^{-1}$, and the median SFR is about 
$184\,M_\odot$\, yr$^{-1}$, which is similar to the values in Daddi \etal 
(2004) (with a typical SFR of $\sim 200\,M_{\odot}$\,yr$^{-1}$) and in 
Kong \etal (2006) (with a median SFR of $\sim190$ $M_\odot$\,yr$^{-1}$).

\subsection{The SFR$-M_*$ correlation}

\begin{figure*}
\centering
\includegraphics[angle=-90,width=0.90\textwidth]{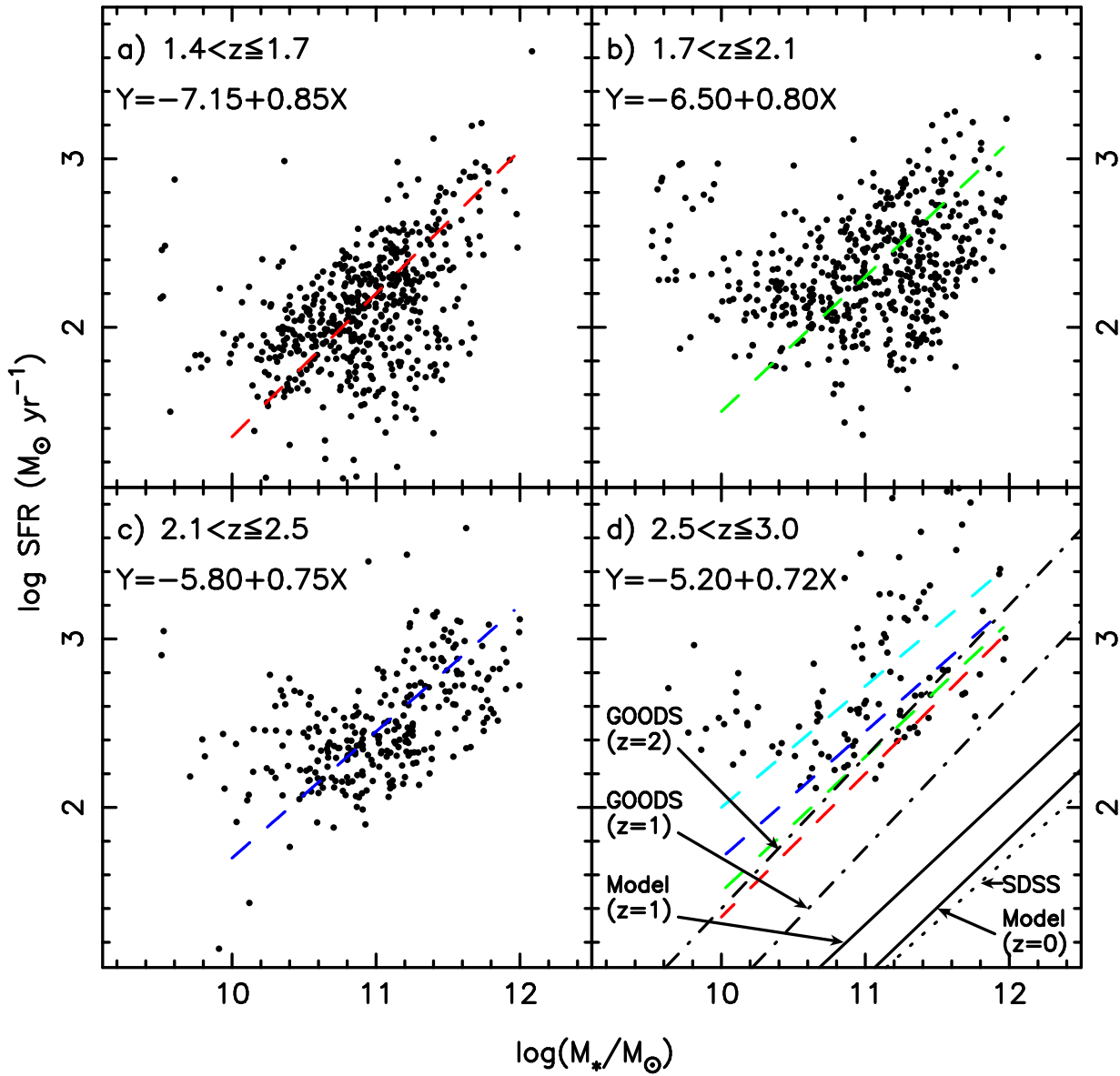}
\caption{
Relationship between stellar mass with SFR, for $z\sim2$ 
star-forming galaxies in 
our sample for each redshift bin. Dashed lines and the equations (
here $Y$ = log SFR, $X$ = log $(M_*/M_\odot)$) in each panel show the 
linear fit for galaxies with $M_*>10^{10} M_\odot$.
The dot-dashed lines are the $z=2$ and $z=1$ correlations from 
Daddi \etal (2007) and Elbaz \etal (2007).
The solid lines show a prediction for $z=0$ and $z=1$ from the 
Millennium simulations (Kitzbichler \& White 2007).
The dashed line is the $z=0.1$ correlation from SDSS (Brinchmann 
\etal 2004). 
\label{fig:sfrm}}
\end{figure*}

The existence of a correlation between star formation and stellar mass 
of galaxies at different redshifts ($0<z<2$) has been reported recently
(Daddi \etal 2007; Elbaz \etal 2007; Pannella \etal 2009). 
To gain further insights into the nature of star formation at $1.4<z<3.0$,
we show the SFR as a function of $M_*$ in the four redshift bins in 
Figure~\ref{fig:sfrm}. To examine in detail the stellar mass dependence 
of the SFR, we plot a good-fit line (dashed line) for galaxies 
with $M_*>10^{10} M_\odot$ in each redshift panel. From this figure, 
we find a positive correlation between the SFR and stellar mass 
of \sgzks\ in the all four redshift bins.

In Figure~\ref{fig:sfrm}(d), we overplot the correlation between the SFR
and stellar mass for $z\sim 0.1$ galaxies in the Sloan Digital Sky Survey (SDSS; Brinchmann \etal 
2004), $z\sim1$ and $z\sim2$ galaxies in the GOODS (Elbaz \etal 2007; 
Daddi \etal 2007), and $z\sim0$ and $z\sim1$ from Millennium model 
(Kitzbichler \& White 2007). Figure~\ref{fig:sfrm}(d) shows the slopes 
between the SFR and stellar mass of galaxies at different redshifts are 
very similar. On average, the SFR of galaxies on the SFR$-M_*$ diagram 
increases with redshift at a fixed $M_*$, by a factor about 6 
from $z \sim 0$ to $z \sim 1$, and a factor about 20 from $z \sim 0$ to $z \sim 2$. 
At fixed stellar mass, star-forming galaxies were much more active on 
average in the past. This is most likely due to a larger abundance of 
gas, depleted with passing of time (Daddi \etal 2007). 

The Millennium model overlaps very well with the SDSS at $z \sim 0.1$ 
and predicts a slope very similar to the observed trend at different
redshift bins. For $z \sim 1$, however, we find that at fixed stellar masses 
the model predicts an increase of the SFR much below the observed trend.
It seems that a major change required for models would be to increase 
the star formation efficiency (and thus the typical SFR) at all 
masses for star-forming galaxies at redshifts $z>1$.

\subsection{The sSFR$-M_*$ correlation}

\begin{figure*}
\centering
\includegraphics[angle=-90,width=0.90\textwidth]{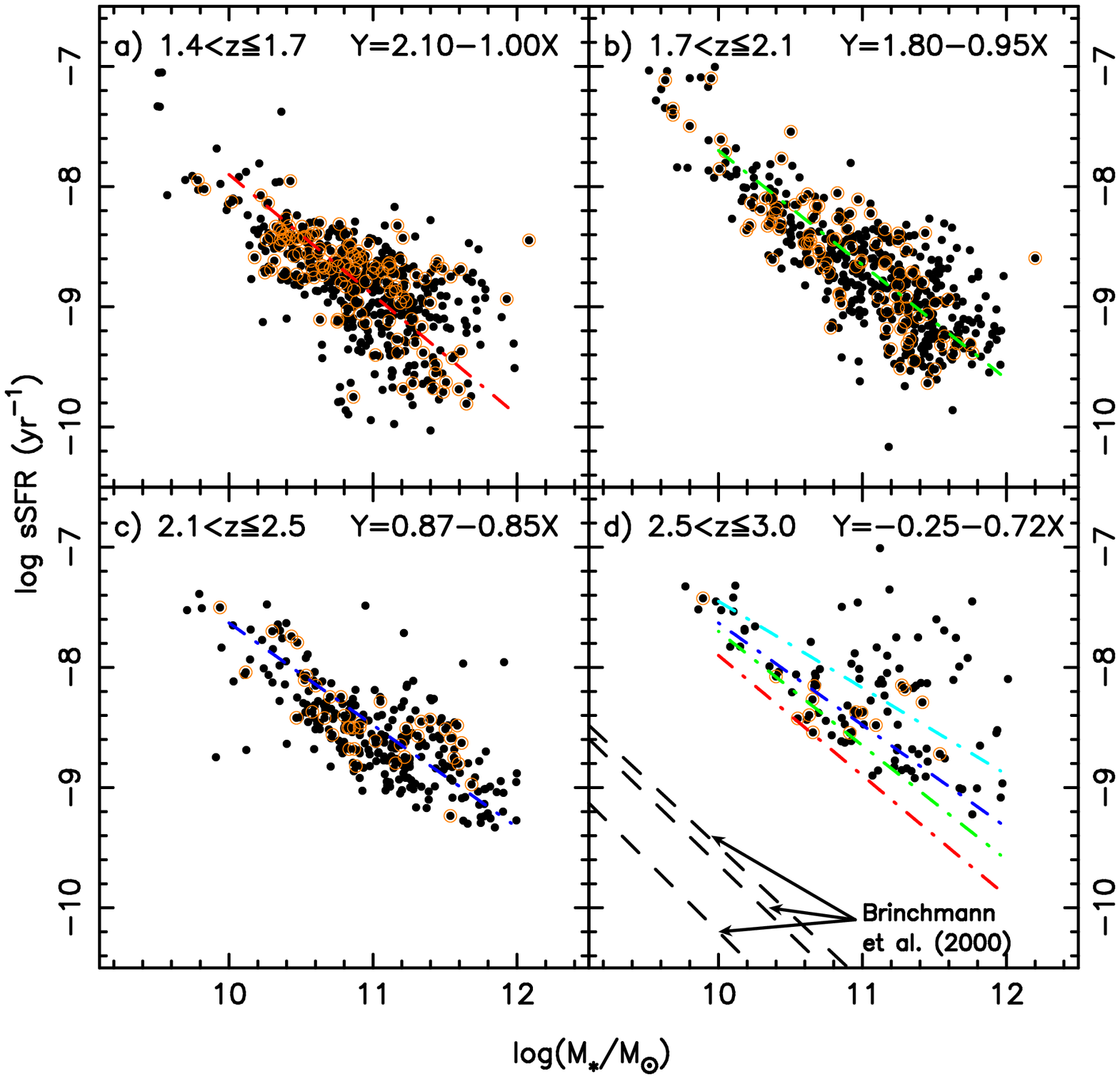}
\caption{
Specific star formation rate (sSFR) as a function of stellar mass 
for \sgzks\ in the EGS. The dashed lines correspond to the best fit to 
the four redshift ranges indicated. The three dashed lines from 
Brinchmann \& Ellis (2000) were overplotted in the bottom-left panel 
(d). The sample of \sgzks\ with open circles from the overlapping area 
between NMBS and AEGIS.
\label{fig:ssfrm}}
\end{figure*}

sSFR is defined as the ratio of the 
current SFR to the current stellar mass, sSFR=SFR$/M_*$. 
Thus, higher values of the sSFR indicate that a larger fraction of 
stars were formed recently, and it can be used to characterize 
the star formation history of galaxies.

Figure~\ref{fig:ssfrm} shows the sSFR as a
function of stellar mass for the \sgzks\ in our sample, with the same
four redshift bins as in Figure~\ref{fig:sfrm}. 
The dashed lines in each panel represent a good fit between the sSFR
 and $M_*$ for galaxies with $M_*>10^{10} M_\odot$. 
In Figure~\ref{fig:ssfrm}(d), we overplot the correlation between the 
sSFR and stellar mass for $0.2<z<0.5$, $0.5<z<0.75$, and $0.75<z<1.0$ 
galaxies in Brinchmann \& Ellis (2000).

From this figure, we find that the general shape of sSFR$-M_*$ correlation
is in good agreement with a similar study for $z<1.0$ galaxies 
(Brinchmann \& Ellis 2000), and the line of maximum sSFR runs parallel 
to lines of constant SFR. 
Furthermore, this sSFR is generally increasing with increasing redshift 
for all stellar masses, suggesting that star formation contributes more
to the growth of low-mass galaxies than to the growth of high-mass galaxies.
Finally, the most massive galaxies have the lowest sSFR at redshift 
$z<2$; they formed their stars earlier and more rapidly than their low-mass 
counterparts. This is in agreement with the downsizing scenario.

\section{Morphology}

The morphology of a galaxy reflects its dynamical history and evolution. 
For instance, different Hubble types are associated with different star
formation histories and different patterns of motion of stars and gas.
Furthermore, galaxy morphology correlates with a range of physical 
properties in galaxies, such as stellar mass, luminosity, size, velocity
dispersion, concentration, and, particularly, color; this suggests
that morphology is crucial in our understanding of the formation
and evolution of galaxies (Fang \etal 2009; Kong \etal 2009).

With the advent of the new WFC3 on board {\it HST}, 
with its vastly improved sensitivity and resolution compared to NICMOS, 
it has become possible to analyze the rest-frame optical structure of $z\sim2$
galaxies with an unprecedented level of detail. To describe clearly 
the morphologies of both \sgzks\ and \pgzks, we have performed nonparametric 
measures of galaxy morphology by using data from {\it HST} WFC3/H(F160W) imaging 
(Grogin \etal 2011; Koekemoer \etal 2011), such as Gini coefficient 
(the relative distribution of the galaxy pixel flux values, or $G$) 
and $M_{20}$ (the second-order moment of the brightest 20\% of 
the galaxy's flux). Early-type galaxies have higher $G$ and lower $M_{20}$,
while late-type galaxies have lower $G$ and higher $M_{20}$.
Combining $G$ and $M_{20}$ permits classifying
different galaxy population effectively (Lotz \etal 2006; Kong \etal 2009).

\begin{figure}
\centering
\includegraphics[angle=-90,width=0.95\columnwidth]{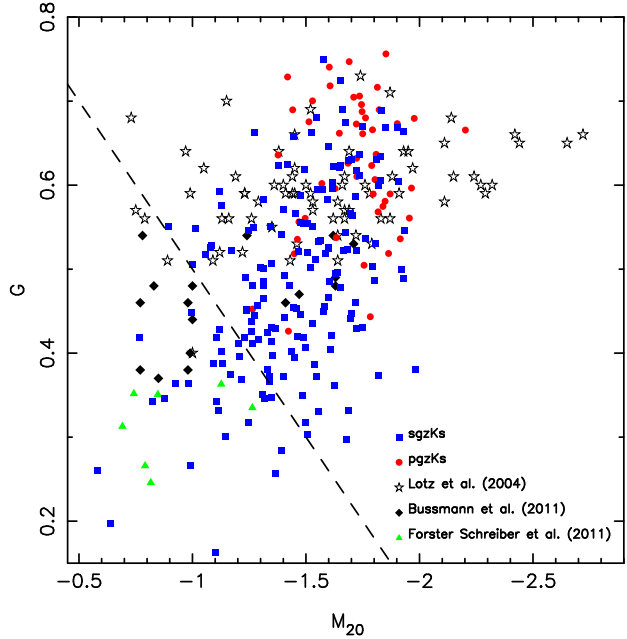}
\caption{$M_{20}$ vs. Gini coefficient ($G$) for sgzKs, pgzKs, 
$z\sim 2$ ULIRGs and local ULIRGs (empty stars; Lotz \etal 2004). 
The dashed line is defined as $G = 0.4M_{20} + 0.9$ (Bussmann \etal 2011). 
Filled blue squares and red circles represent the morphological distribution of
173 \sgzks\ and 50 \pgzks\ in our sample, respectively, which are matched in 
{\it HST}/WFC3 $H$-band. The starburst-dominated sources at $z\sim2$ from 
F\"orster Schreiber \etal (2011) and Bussmann \etal (2011) are shown 
as filled triangles and diamonds, respectively, with $G$ and $M_{20}$ derived 
from {\it HST} NICMOS/F160W images. 
\label{fig:mor}}
\end{figure}

In Figure~\ref{fig:mor}, the blue squares and 
red circles represent the morphological distribution of 173
sgzKs and 50 pgzKs (these sources have $H$-band counterparts) in the rest-frame 
wavelength $\sim 5300$\AA, respectively. The distribution of gzKs is
very similar to that of local galaxies (E-Sd) in Lotz \etal 2004, with
pgzKs showing high $G$ and low $M_{20}$, and sgzKs with lower $G$ and
higher $M_{20}$ values. This figure clearly illustrates the morphological
variety present at $z\sim2$. The median ($G$, $M_{20}$) values for
sgzKs are (0.49, --1.48) in the rest-frame optical band, while pgzKs are 
(0.64, --1.74). A sample of 73 local ULIRGs ($z<0.2$) is also shown in this diagram 
(Lotz \etal 2004), indicated as empty stars, using data from {\it HST} WFPC 2/F814W 
imaging (Borne \etal 2000). The solid diamonds and triangles in
Figure~\ref{fig:mor} represent the starburst-dominated sources
at $z\sim2$ from Bussmann \etal (2011) and F\"orster Schreiber \etal (2011), 
respectively; their $G$ and $M_{20}$ values were derived from NICMOS/F160W images,
corresponding to the rest-frame optical wavelength. The dashed line
($G = 0.4M_{20} + 0.9$) in this figure is drawn qualitatively
at rest-frame optical band ($\sim 5300$\AA; Bussmann \etal 2011).
Galaxies with $G < 0.4M_{20} + 0.9$ have diffuse structures or multiple 
bright nuclei in appearance. Objects with $G > 0.4M_{20} + 0.9$
and $-1.6 < M_{20}$ can be irregular, often with a bright nucleus 
with tidal tail or fainter knots. As for sources with 
$G > 0.4M_{20} + 0.9$ and $M_{20} < -1.6$,
they are relatively smooth with a single nucleus. From 
Figure~\ref{fig:mor}, we find that star-forming galaxies at 
high redshift not only have diffuse structures, similar to
starburst-dominated ULIRGs at $z\sim2$ with lower $G$ and 
higher $M_{20}$ values, but also have single-object morphologies
(higher $G$ and lower $M_{20}$). Moreover, the distribution
of $M_{20}$ of sgzKs is mainly located in the range from --1.0
to --1.8; these features of morphologies are consistent with
expectations from simulations of major mergers during the
beginning and end stages, respectively, of the phase of the
merger when the SFR peaks and begins to turn over (Lotz \etal 2008). 

\begin{figure}
\centering
\includegraphics[angle=0,width=\columnwidth]{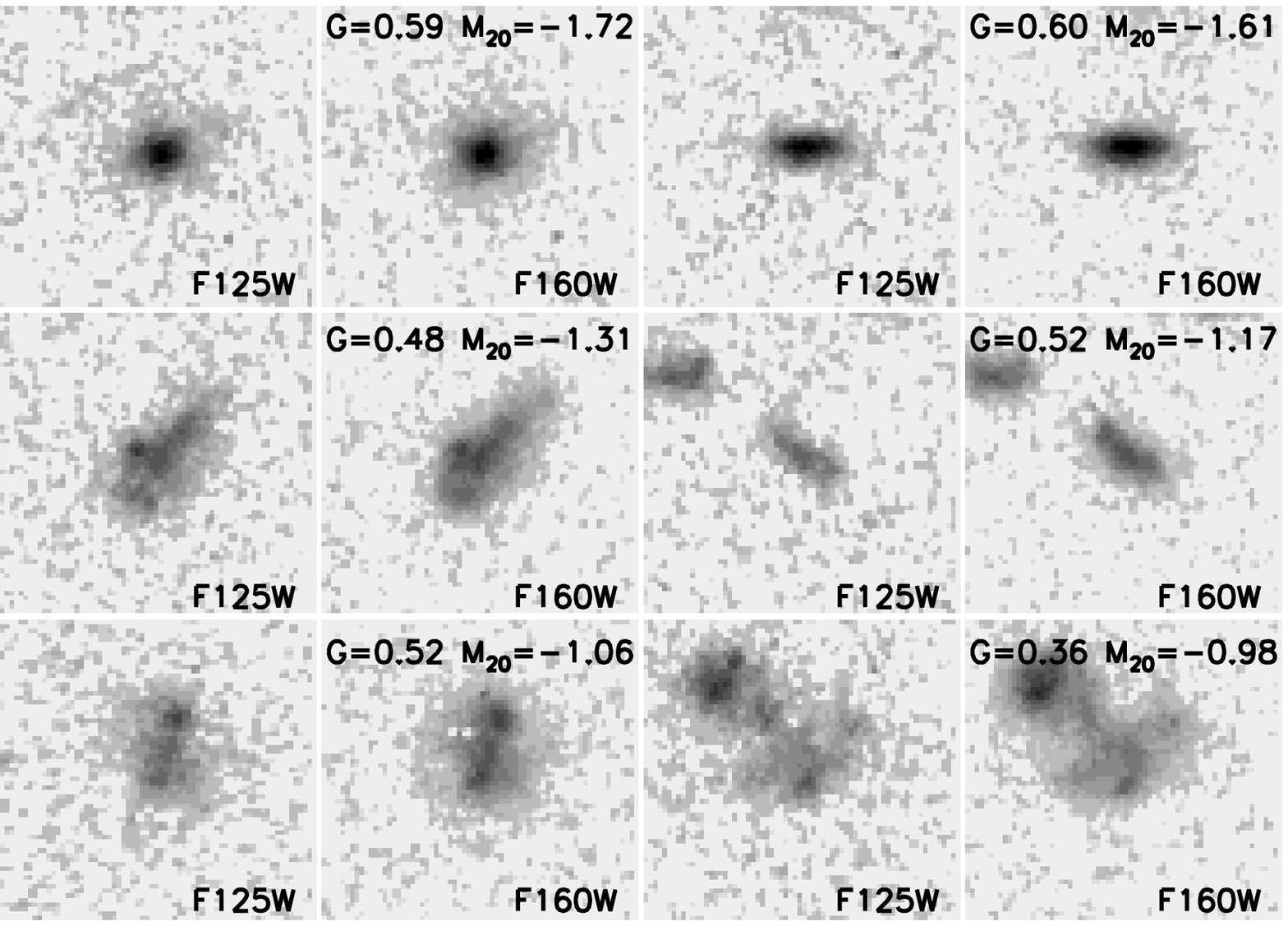}
\caption{{\it HST}/WFC3 $J$- and $H$-band images of six sgzKs in the EGS.
Top: $M_{20} < -1.60$, middle: $-1.60< M_{20} < -1.10$, bottom:
$M_{20} > -1.10$. Values of $G$ and $M_{\rm 20}$ are shown in 
each $H$-band image. All images are in negative gray scale and 
are $4\times4$~arcsec$^2$ in size.
\label{fig:mors}}
\end{figure}

\begin{figure}
\centering
\includegraphics[angle=0,width=\columnwidth]{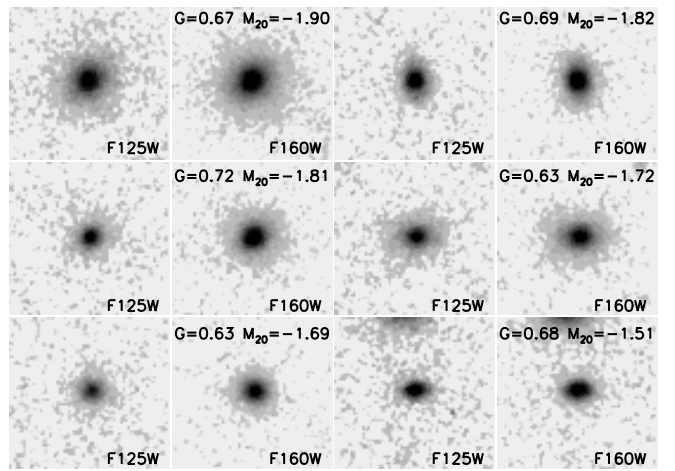}
\caption{{\it HST}/WFC3 $J$- and $H$-band images of six pgzKs in the EGS.
Top: $M_{20} < -1.81$, middle: $-1.81< M_{20} < -1.70$, bottom:
$M_{20} > -1.70$. Values of $G$ and $M_{\rm 20}$ are shown in 
each $H$-band image. All images are 
in negative gray scale and are $4\times4$~arcsec$^2$ in size.
\label{fig:morp}}
\end{figure}

Figures~\ref{fig:mors} and~\ref{fig:morp} show the {\it HST}/WFC3 $J$- and $H$-band
stamp images for a subset of six sgzKs and six pgzKs, respectively. Their rest-frame 
optical morphologies for sgzKs are very diversified including string-like, 
extended/diffused, and even early-type spiral morphologies, while pgzKs are
relatively smooth and compact, implying that there are different formation 
process for these galaxies. Therefore, we conclude that the population of
Hubble sequence galaxies roughly matches that of the peculiars sometime 
between $z=1.5$--2.

\section{Fraction of AGNs}\label{sec:agn}

Mid-IR photometry has been proven to be a robust and efficient tool 
to select AGNs without prior information (e.g., Lacy \etal 2004; 
Stern \etal 2005; Donley \etal 2008), as their properties at these 
wavelengths are typically very different from those of stars and 
galaxies. There have been several studies of AGN selection using mid-IR 
color or other IR properties. In this section, two different AGN 
selection methods are employed, to identify AGNs in our gzK sample.
The first one is based on the IRAC band color criteria of Stern \etal 
(2005). The second method is based on Mid-IR spectral index (Barmby 
\etal2006; Park \etal 2010).

\subsection{Mid-IR colors}

\begin{figure}
\centering
\includegraphics[angle=-90,width=0.95\columnwidth]{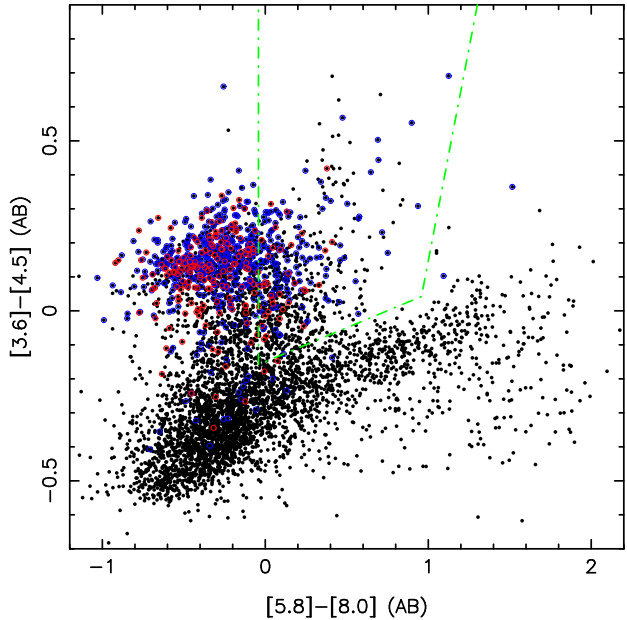}
\caption{
IRAC band color--color diagram of galaxies in the EGS field.
Dot-dashed lines indicate the boundaries of the AGN selection 
wedges from Stern \etal (2005). Small points represent all 
galaxies, \sgzks\ are denoted by blue open circles, and \pgzks\ are
denoted by red open circles. 
\label{fig:mid}}
\end{figure}

Figure~\ref{fig:mid} shows the mid-IR color space of the EGS survey
 along with the Stern \etal (2005) AGN selection criterion (green 
dot-dashed lines). Small dots represent the distribution of 
Mid-IR color for sources detected in all four IRAC channels in 
the AEGIS field; they have very similar 
[3.6] -- [4.5] colors, as this is dominated by their stellar emission.
Six hundred and twelve \sgzks\ in our sample are detected in all the four IRAC channels 
and are plotted as blue open circles. One hundred and twenty-two of them can be selected as AGN
candidates by the mid-IR selection criterion of Stern \etal (2005).
Two hundred and sixteen \pgzks\ in our sample are detected in all the four IRAC channels 
and plotted as red open circles. Thirty-five of them are AGN candidates. 
Therefore, using the mid-IR AGN selection criterion, $\sim20\%$ of 
\sgzks\ and $\sim16\%$ of \pgzks\ are classified as AGNs, 
respectively.

The AGN selection criterion from Stern \etal (2005) were based on the
 spectroscopic sample of the AGN and Galaxy Evolution Survey, 
most of which have redshift $z < 0.6$. Their mid-IR color criterion 
is reliable for classifying AGNs and galaxies at low redshift.
In contrast, observations and templates suggest that a high degree 
of stellar contamination is unavoidable when the color selection 
technique is applied to deeper samples (Barmby \etal 2006; 
Donley \etal 2008; Georgantopoulos \etal 2008; Assef \etal 2010).
The redshift of \sgzks\ and \pgzks\ in our sample is $z \sim 2$;
with much deeper IRAC data, we will classify AGNs from normal 
galaxies with the mid-IR spectral index method in the next 
subsection.
 
\subsection{Mid-IR spectral index}

The slope of a galaxy SED in the infrared can be characterized
by a power-law behavior of flux density with frequency 
$f_\nu\propto \nu^\alpha$. AGN SEDs often follow a negative-sloping 
(red) power-law, which may arise from either thermal or non-thermal 
emission originating near the central region. 
In contrast, stellar-dominated sources at redshift $z\lsim2$ 
generally exhibit positive (blue) IRAC power law emission.
For this reason, mid-IR power-law $\alpha$ of galaxies have been 
proposed and used as classification criteria for AGNs and normal
galaxies, with a small level of galaxy contamination.

In this paper, the IRAC fluxes for each gzKs, covering the 
3.6--8.0 $\mu$m range, are fitted with a power law, $\alpha$.
The $\alpha$ value is only accepted if the $\chi^2$ probability
fit is $P_{\chi^2}>0.1$. Galaxies fitted well with a power-law SED 
are further classified as red or blue. A limit of $\alpha \lsim-0.5$, 
following  Donley \etal (2008) and  Park \etal (2010), is chosen to 
classify gzKs as AGNs. The criterion reveals that the fraction of AGNs 
in our sample is $\sim$ 10\%. In the meantime, we also check the X-ray 
luminosities of these AGN candidates (if they have X-ray counterparts), 
and the result shows that most of these objects 
have $L_{\rm 0.5-10~keV}>10^{41}\,\rm erg\,s^{-1}$.      
 
\section{Summary}\label{sec:sum}

In this paper, we have described the construction of a sample of \gzks\
(correspond to the BzKs of Daddi \etal2004) within the AEGIS. With
this field's unique combination of area and depth at almost
every waveband observable, we analyze the photometric redshifts, number
counts, SFRs, sSFRs, stellar masses, morphologies, and the fraction of AGNs 
of \gzks\ in the AEGIS. Our main conclusions are as follows.

1. We identify a sample of 2031 \gzks\, including 1609 star-forming galaxies, with
$gzK = (z-K)_{\rm AB}-1.4(g-z)_{\rm AB}\geq0.2$ and 422 passively evolving galaxies
with $gzK<0.2$ and $(z-K)_{\rm AB}>2.7$, to a limit $K_{\rm AB}<22$. 
The surface density of \sgzks+\pgzks\ is $\sim 1.28$ arcmin$^{-2}$. Of the sgzK galaxies,
83.2\% lie in the range $1.4\leq z\leq2.5$, while 8.5\% are at $z<1.4$ and 8.3\%
at $z>2.5$. For pgzK galaxies, there is a considerable fraction (12.8\%) of 
objects at low redshift ($z<1.0$). The $z_{\rm p}$ median values of both 
\sgzks\ and \pgzks\ are 1.8 and 1.5, respectively.

2. For a sample selected at $K_{\rm AB} < 22.0$, 2989 spectroscopic redshift 
of field galaxies (including 35 LBGs at $z\sim3$ and 14 ULIRGs at $z\sim2$),
we reach an average redshift accuracy of $\delta z/(1+z_{\rm s}) = -0.014$ 
with the normalized median absolute deviation ($\sigma_{\rm NMAD}$) of 0.032. 
Only 4.3\% galaxies have $(\zph-\zsp)/(1+\zsp)>0.1$. We also
compare our photometric redshifts with NMBS photometric redshifts in the 
AEGIS field; they are in good agreement.

3. Similar to the findings in other literatures, our counts of passively 
evolving galaxies turn over at $K_{\rm AB} \sim 21.0$ and both the number of 
faint and bright sources (including \sgzks\ and \pgzks) in our catalogs 
exceed the predictions of a recent semi-analytic model of galaxy formation, 
implying that a more successful model must have to explain this diversity 
in the future.

4. Based on the reddening-corrected UV luminosities and SED fitting of
the multi-band photometry, we find that {\it gzK}-selected star-forming galaxies have 
a median SFR and stellar mass of $\sim 184\,M_\odot$\,yr$^{-1}$  
and $\sim 8.8\times10^{10}~M_\odot$, respectively. Moreover, we also find that the SFR 
and sSFR of \sgzks\ increases with redshift at all masses, suggesting that 
star-forming galaxies were much more active on average in the past.
Comparing to the predictions of the Millennium model, there is a major change
required for models in order to increase the star formation efficiency at all 
masses for star-forming galaxies at redshifts $z>1$.

5. We also study the morphological properties of gzKs in our sample,
employing data from {\it HST} WFC3/F160W imaging within the EGS field. We 
derived the median ($G$, $M_{20}$) values for sgzKs and pgzKs are (0.49, --1.48) 
and (0.64, --1.74) in the rest-frame optical band, respectively. 
Moreover, we find that morphologies of $z\sim2$ galaxies in our sample are 
complex and varied: from compact, apparently early-type galaxies to large 
star-forming systems superficially similar to pure disk or irregular galaxies,
implying that there are different formation process for these galaxies.
We conclude that the population of Hubble sequence galaxies roughly matches 
that of the peculiars sometime between $z=$1.5--2.
%Almost all of the physical properties of sgzKs such as stellar mass, SFR, and
%morphology, indicating that gzK selection might probe both
%massive star-bursting galaxies and massive, extreme, dusty
%starbursts.

6. For the fraction of AGNs in the \gzks, we compared the power-law selection 
technique with mid-infrared color selection, and we find that 82 sources 
with $\alpha\leq-0.5$, corresponding to the fraction of AGNs, is $\sim$ 10\% in 
our sample of 828 gzK galaxies with four IRAC bands. In the meantime, we also 
check the X-ray luminosities of these sources (if they have X-ray counterparts);
the result shows that most of these objects 
have $L_{\rm 0.5-10~keV}>10^{41}\,\rm erg\,s^{-1}$. 

\acknowledgments
We thank the anonymous referee for valuable comments and suggestions 
that have improved the paper.
We thank J.-S. Huang and H. J. McCracken for helpful suggestions and 
discussions.
This study makes use of data from AEGIS, a multi-wavelength sky survey 
conducted with the {\it Chandra}, {\it GALEX}, {\it Hubble}, Keck, CFHT, MMT, Subaru, 
Palomar, {\it Spitzer}, VLA, and other telescopes and supported in part by 
the NSF, NASA, and the STFC.
This work is also based on observations taken by the CANDELS 
Multi-Cycle Treasury Program with the NASA/ESA {\it HST}, which is operated 
by the Association of Universities for Research in Astronomy, Inc., 
under NASA contract NAS5-26555. 
The work is supported by the National Natural Science Foundation of 
China (NSFC, No. 10873012), the Open Research Program of Key Laboratory 
for the Structure and Evolution of Celestial Objects, CAS, and Chinese 
Universities Scientific Fund (CUSF).

\end{document}